\documentstyle[prl,aps,multicol]{revtex}

\renewcommand{\narrowtext}{\begin{multicols}{2} \global\columnwidth20.5pc}
\renewcommand{\widetext}{\end{multicols} \global\columnwidth42.5pc}
\begin{document}
\draft
\title{Higher order parametric level statistics in disordered systems}
\author{E. Kanzieper$^1$ and V. Freilikher$^2$}
\address{$^1$The Abdus Salam International Centre for Theoretical Physics, P.O.B. 586, 34100 Trieste, Italy\\
$^2$The Jack and Pearl Resnick Institute of Advanced Technology,\\
Department of Physics, Bar--Ilan University, 52900 Ramat--Gan, Israel}
\maketitle

\begin{abstract}
Higher order parametric level correlations in disordered systems with broken
time--reversal symmetry are studied by mapping the problem onto a model of
coupled Hermitian random matrices. Closed analytical expression is derived
for parametric density--density correlation function which corresponds to a
perturbation of disordered system by a multicomponent flux.
\end{abstract}


\narrowtext

Parametric level statistics reflects the response of the spectrum $\left\{
E_n\right\} $ of complex chaotic systems to an external perturbation. A few
years ago it was shown \cite{SzA-1993,SA-1993,SLA-1993} that a system whose
spectrum follows closely the universal fluctuations predicted by the random
matrix theory \cite{M-1991} should also exhibit a universal parametric
behavior. This conclusion was reached by analyzing the dimensionless
autocorrelator of level velocities of electron in a disordered metallic
sample with a ring topology, enclosing a magnetic flux $\varphi $. A
diagrammatic perturbation technique was used in the range of fluxes $%
g^{-1/2}\ll \varphi \ll 1$ \cite{SzA-1993}, while the opposite limit, $%
\varphi <g^{-1/2}$ \cite{SA-1993}, has been treated within the framework of
the supersymmetry formalism \cite{E-1983,VWZ-1985}. [Here $g\gg 1$ is the
dimensionless conductance]. In this particular problem, the parametric
correlations take a universal form involving the rescaled parameter $%
X^2=4\pi g\varphi ^2$, with $g=E_c/\Delta $, the ratio of the Thouless
energy and the mean level spacing. Numerical simulations have supported the
point that the universal character of parametric level statistics extends to
a wider class of chaotic systems without disorder [chaotic billiards] whose
Hamiltonian depends on some external parameter $x$. In such systems, the
spectral fluctuations taken at different values of $x$ become
system--independent after rescaling, $x\rightarrow X$, which involves solely
the `generalized' dimensionless conductance $g=\left( 4\pi \right)
^{-1}\Delta ^{-2}\left\langle {}\right. \left( \partial E_n\left( x\right)
/\partial x\right) ^2\left. {}\right\rangle $. Along with the diagrammatic
technique and the supersymmetry formalism, the parametric correlations have
been studied in detail within the model of Brownian motion \cite
{NS-1993,BR-1994} and in the semiclassical limit \cite{BK-1994,ALM-1997}.

A further burst of activity in the field occurred after it was realized \cite
{SLA-1993,NS-1993,SLA-1994,TSA-1995,ZH-1995,NP} that the problem of
parametric level correlations is identical to the ground state dynamics of
the integrable many--body quantum model known as Calogero--Sutherland--Moser
[CSM] system. This gave new information about CSM space--time $\left( r,\tau
\right) $ correlation functions that can be obtained from parametric
density--density correlators $\left\langle {}\right. \nu (E,0)\nu (E^{\prime
},\varphi )\left. {}\right\rangle $ involving only two different external
parameters by mapping \cite{SLA-1993} $X^2\rightarrow -2i\tau $, $E/\Delta
\rightarrow r$. For the more general situation of higher order correlation
functions the connection between CSM fermions and quantum chaotic systems
has been established \cite{NP} as well by using the supersymmetry technique,
however it has not led to any explicit analytical results beyond the
two--point correlators due to enormous increase of number of entries in the
supermatrix fields, thereby making any explicit calculations in that
approach impossible. Extensions to higher order statistics can be performed
by using an involved method of differential equations for quantum
correlation functions proposed in the much earlier work \cite{IIKS-1990}.

In the present paper we address the issue of higher order parametric level
statistics within the framework of the random matrix theory, by appealing to
the model of coupled Hermitian random matrices \cite{M-1997}. The latter
enables us to provide a complete information about parametric correlations
of single electron level densities in the presence of the multicomponent
flux perturbing a disordered system, characterized by a dimensionless
conductance $g\gg 1$. To the best of our knowledge, this is the first
detailed study of higher order parametric level statistics in disordered
systems which adopts the conventional language of the random matrix theory.

In what follows we consider a weakly disordered system fallen in the
universal (metallic) regime, $g\gg 1$, which is known \cite{E-1983} to be
modelled by invariant ensembles of large random matrices. Assuming that the
time reversal symmetry is completely broken (unitary symmetry), one can
statistically describe an unperturbed single electron spectrum by Gaussian
Unitary Ensemble [GUE] of large $N\times N$ random matrices $H_0$
distributed in accordance with the probability density ${\cal P}\left[
H_0\right] \propto \exp \left\{ -%
\mathop{\rm Tr}
H_0^2\right\} $. Such a distribution ${\cal P}\left[ H_0\right] $ induces
the energy scale $\Delta $ being the mean level spacing, $\Delta =\pi \left(
2N\right) ^{-1/2}$. Let us now apply a Gaussian perturbation consisting of $d
$ components $\overrightarrow{{\bf \varphi }}_d=\left( \phi _1,\ldots ,\phi
_d\right) $ which does not change the global unitary symmetry, and which
drives the Hamiltonian $H_0$ to $H=H_0+\sum_{k=1}^d\phi _kH_k$, with
matrices $H_k$ drawn from GUE: ${\cal P}\left[ H_k\right] \propto \exp
\left\{ -%
\mathop{\rm Tr}
H_k^2\right\} $ for $k=1,\ldots ,d.$ This choice corresponds to the equal
`strength' of each component of the `vector' perturbation $\overrightarrow{%
{\bf \varphi }}_d$ since the average $\left\langle (H_k)_{\mu \nu
}(H_k)_{\mu ^{\prime }\nu ^{\prime }}\right\rangle $ is independent of the
index $k$. The quantity which provides the most detailed information about
parametric correlations in the case of the multicomponent perturbation $%
\overrightarrow{{\bf \varphi }}_d$ is the correlator of level densities $\nu
(E,\overrightarrow{{\bf \varphi }}_\sigma )=%
\mathop{\rm Tr}
\delta (E-H_0-\sum_{k=1}^\sigma \phi _kH_k)$ taken at both different values
of energy $E$ and of $\sigma $. For this reason, we will concentrate on the
dimensionless multipoint correlator\widetext
\begin{equation}
k_{p_0,\ldots ,p_d}(\{\omega ^{\left( 0\right) }\},\overrightarrow{0};\ldots
;\{\omega ^{\left( d\right) }\},\overrightarrow{{\bf \varphi }}_d)=\Delta
^m\left\langle \prod_{i_0=1}^{p_0}\nu (E+\omega _{i_0}^{\left( 0\right) },%
\overrightarrow{0})\prod_{i_1=1}^{p_1}\nu (E+\omega _{i_1}^{\left( 1\right)
},\overrightarrow{{\bf \varphi }}_1)\,\ldots \prod_{i_d=1}^{p_d}\nu
(E+\omega _{i_d}^{\left( d\right) },\overrightarrow{{\bf \varphi }}%
_d)\right\rangle ,  \label{def}
\end{equation}
where $m=p_0+\ldots +p_d$, and the angular brackets stand for averaging over
ensembles of Hermitian matrices $H_k$ with $k=0,\ldots ,d.$ Equation (\ref
{def}) can be rewritten as a $(d+1)$ multiple matrix integral over matrices $%
\widetilde{H}_0=H_0$ and $\widetilde{H}_{\sigma >0}=H_0+\sum_{k=1}^\sigma
\phi _kH_k$, 
\begin{equation}
k_{p_0,\ldots ,p_d}\propto \int d\widetilde{H}_0\,\ldots \int d\widetilde{H}%
_d\prod_{\sigma =0}^d\prod_{i_\sigma =1}^{p_\sigma }%
\mathop{\rm Tr}
\delta \left( E+\omega _{i_\sigma }^{\left( \sigma \right) }-\widetilde{H}%
_\sigma \right) \exp \left\{ -%
\mathop{\rm Tr}
\left[ \sum_{\alpha =0}^d\left( \phi _\alpha ^{-2}+\phi _{\alpha
+1}^{-2}\right) \widetilde{H}_\alpha ^2-2\sum_{\alpha =0}^{d-1}\phi _{\alpha
+1}^{-2}\widetilde{H}_\alpha \widetilde{H}_{\alpha +1}\right] \right\} ,
\label{mi}
\end{equation}
\narrowtext
\noindent with $\phi _0=1$ and $\phi _{d+1}=\infty .$ [This convention is
relaxed everywhere below Eq. (\ref{pol})]. We notice that the strengths $%
\phi _k$ $\left( k=1,\ldots ,d\right) $ of the perturbation are supposed to
be small, $\phi _k\ll 1$. This is justified in the thermodynamic limit $%
N\rightarrow \infty $, since for Gaussian perturbation accepted above, $\phi
_k$ are known to scale with $N$ as $\phi _k=\pi N^{-1/2}X_k$, with $X_k$
being the set of dimensionless parameters of order unity \cite{RemarkP}.

Our crucial observation is that Eq. (\ref{mi}) can be interpreted as a
density--density correlator in the effective model of $(d+1)$ Hermitian
random matrices coupled in a chain: Each matrix $\widetilde{H}_\alpha $ is
represented by a point, and two adjacent matrices $\widetilde{H}_\alpha $
and $\widetilde{H}_{\alpha +1}$ are joined by a line if the coupling of the
type $\exp \{c_\alpha 
\mathop{\rm Tr}
\widetilde{H}_\alpha \widetilde{H}_{\alpha +1}\}$ is present in Eq. (\ref{mi}%
). In this situation, the joint probability density of eigenvalues of all
the matrices in the chain can be deduced through the Itzykson--Zuber
integral \cite{IZ-1980} making the model of random Hermitian matrices
coupled in a chain to be a completely solvable. In accordance
with the Eynard--Mehta theorem \cite{M-1997}, the dimensionless correlator $%
k_{p_0,\ldots ,p_d}$ can be represented as a determinant of the $m\times m$
block matrix, $m=p_0+\ldots +p_d$, consisting of $(d+1)\times (d+1)$
rectangular submatrices $K_{\alpha ,\beta }$ with $\alpha ,\beta =1,\ldots
,\left( d+1\right) $, each of them having $p_{\alpha -1}\times p_{\beta -1}$
entries \cite{RemarkM},\widetext
\begin{equation}
k_{p_0,\ldots ,p_d}=%
\mathop{\rm Det}
\left( 
\begin{array}{cccc}
\fbox{$K_{1,1}\left( \omega _{i_0}^{\left( 0\right) },\omega _{j_0}^{\left(
0\right) }\right) $}_{p_0\times p_0} & \fbox{$K_{1,2}\left( \omega
_{i_0}^{\left( 0\right) },\omega _{j_1}^{\left( 1\right) }\right) $}%
_{p_0\times p_1} & \cdots  & \fbox{$K_{1,d+1}\left( \omega _{i_0}^{\left(
0\right) },\omega _{j_d}^{\left( d\right) }\right) $}_{p_0\times p_d} \\ 
\fbox{$K_{2,1}\left( \omega _{i_1}^{\left( 1\right) },\omega _{j_0}^{\left(
0\right) }\right) $}_{p_1\times p_0} & \fbox{$K_{2,2}\left( \omega
_{i_1}^{\left( 1\right) },\omega _{j_1}^{\left( 1\right) }\right) $}%
_{p_1\times p_1} & \cdots  & \fbox{$K_{2,d+1}\left( \omega _{i_1}^{\left(
1\right) },\omega _{j_d}^{\left( d\right) }\right) $}_{p_1\times p_d} \\ 
\vdots  & \vdots  & \ddots  & \vdots  \\ 
\fbox{$K_{d+1,1}\left( \omega _{i_d}^{\left( d\right) },\omega
_{j_0}^{\left( 0\right) }\right) $}_{p_d\times p_0} & \fbox{$K_{d+1,2}\left(
\omega _{i_d}^{\left( d\right) },\omega _{j_1}^{\left( 1\right) }\right) $}%
_{p_d\times p_1} & \cdots  & \fbox{$K_{d+1,d+1}\left( \omega _{i_d}^{\left(
d\right) },\omega _{j_d}^{\left( d\right) }\right) $}_{p_d\times p_d}
\end{array}
\right) .  \label{det}
\end{equation}
\narrowtext
\noindent The matrix kernels $K_{\alpha ,\beta }$ in Eq. (\ref{det}) are 
\begin{equation}
K_{\alpha ,\beta }\left( \xi ,\eta \right) =\Delta \left[ H_{\alpha ,\beta
}\left( \xi ,\eta \right) -E_{\alpha ,\beta }\left( \xi ,\eta \right)
\right] ,  \label{k1}
\end{equation}
where 
\begin{equation}
H_{\alpha ,\beta }\left( \xi ,\eta \right) =\sum_{j=0}^{N-1}\frac 1{h_j}%
Q_{\alpha ,j}\left( \xi \right) P_{\beta ,j}\left( \eta \right) ,  \label{k2}
\end{equation}
and 
\begin{equation}
E_{\alpha ,\beta }\left( \xi ,\eta \right) =\left( w_\alpha *\ldots
*w_{\beta -1}\right) \left( \xi ,\eta \right)   \label{k3}
\end{equation}
for $1\leq \alpha <\beta \leq d+1$; otherwise, $E_{\alpha ,\beta }=0$. Here
the partial weights $w_\alpha $ are 
\begin{mathletters}
\label{wv}
\begin{eqnarray}
w_\alpha (\xi ,\eta ) &=&\exp \{-\frac{V_\alpha (\xi )+V_{\alpha +1}(\eta )}2%
+2\phi _\alpha ^{-2}\xi \eta \},  \label{w} \\
V_\alpha (\xi ) &=&(\phi _{\alpha -1}^{-2}+\phi _\alpha ^{-2})[\delta
_{\alpha ,1}+\delta _{\alpha ,d+1}+1]\xi ^2,  \label{v}
\end{eqnarray}
\end{mathletters}
(compare with the weight of the matrix model, Eq. (\ref{mi})). The notation $%
\left( w_\alpha *\ldots *w_{\beta -1}\right) \left( \xi ,\eta \right) $
stands for the product of the partial weights $w$ integrated over internal
variables of that product. Two sets of orthogonal functions $P_{\alpha ,j}$
and $Q_{\beta ,j}$ entering Eq. (\ref{k2}) are determined recursively 
\begin{mathletters}
\label{sets}
\begin{eqnarray}
P_{\alpha ,j}\left( \xi \right)  &=&\int d\eta P_{\alpha -1,j}\left( \eta
\right) w_{\alpha -1}\left( \eta ,\xi \right) ,  \label{p.set} \\
Q_{\beta ,j}\left( \xi \right)  &=&\int d\eta w_\beta \left( \xi ,\eta
\right) Q_{\beta +1,j}\left( \eta \right) ,  \label{q.set}
\end{eqnarray}
\end{mathletters}
for $2\leq \alpha \leq d+1$ and $1\leq \beta \leq d$; the starting points of
the recursions (\ref{p.set}) and (\ref{q.set}) are the polynomials $%
P_{1,j}=P_j$ and $Q_{d+1,j}=Q_j$ orthogonal with respect to a {\it nonlocal}
weight $W\left( \xi ,\eta \right) =\left( w_1*\ldots *w_d\right) \left( \xi
,\eta \right) $, 
\begin{equation}
\int d\xi \int d\eta P_i\left( \xi \right) W\left( \xi ,\eta \right)
Q_j\left( \eta \right) =h_j\delta _{ij}.  \label{pol}
\end{equation}

Close inspection of the equations above shows that the basic orthogonal
polynomials $P_j$ and $Q_j$ can be expressed in terms of Hermite
polynomials, $P_j(\xi )=H_j(\xi )$, $Q_j(\xi )=H_j(\xi [1+\sum_{k=1}^d\phi
_k^2]^{-1/2})$. Then, step--by--step integrations in Eqs. (\ref{sets}) yield 
\begin{mathletters}
\label{pq-full}
\begin{eqnarray}
P_{\alpha ,j}\left( \xi \right) &=&\frac{\prod_{k=1}^{\alpha -1}\left( \phi
_k\sqrt{\pi }\right) }{\left[ 1+\sum_{k=1}^{\alpha -1}\phi _k^2\right] ^{j/2}%
}%
\mathop{\rm e}
{}^{-F_\alpha \left( \xi \right) }\Phi _j\left( \frac \xi {C_{\alpha -1}}%
\right) ,  \label{p-full} \\
Q_{\alpha ,j}\left( \xi \right) &=&\frac{\prod_{k=\alpha }^d\left( \phi _k%
\sqrt{\pi }\right) }{\left[ 1+\sum_{k=\alpha }^d\phi _k^2\right] ^{j/2}}%
\mathop{\rm e}
{}^{F_\alpha \left( \xi \right) }\Phi _j\left( \frac \xi {C_{\alpha -1}}%
\right) ,  \label{q-full}
\end{eqnarray}
\end{mathletters}
where we have introduced the Hermite functions \cite{RemarkH} $\Phi _j\left(
\xi \right) =\exp \left[ -\xi ^2/2\right] H_j\left( \xi \right) $. Also, we
defined the function 
\begin{equation}
F_\alpha (\xi )=\frac{\xi ^2}2\left[ C_{\alpha -1}^{-2}+\left( \phi _\alpha
^{-2}-\phi _{\alpha -1}^{-2}\right) \right] ,  \label{not}
\end{equation}
and the constant $C_\alpha =[1+\sum_{k=1}^\alpha \phi _k^2]^{1/2}$. [In
order to compactify the formulas, it is agreed from now on that $\phi
_{d+1}=\phi _d$, $\phi _0=\phi _1$, $\sum_{k=\alpha }^{\beta <\alpha }\left(
\ldots \right) =0$, and $\prod_{k=\alpha }^{\beta <\alpha }\left( \ldots
\right) =1$]. One can verify that the orthogonality relation (\ref{pol}) is
satisfied with 
\begin{equation}
h_j=2^jj!\sqrt{\pi }[1+\sum_{k=1}^d\phi _k^2]^{-j/2}\prod_{k=1}^d\left( \phi
_k\sqrt{\pi }\right) ,  \label{nrml}
\end{equation}
so that the first term in Eq. (\ref{k1}) is 
\begin{mathletters}
\label{hh}
\begin{eqnarray}
H_{\alpha ,\alpha }\left( \xi ,\eta \right) &=&%
\mathop{\rm e}
{}^{F_\alpha \left( \xi \right) -F_\alpha \left( \eta \right) }  \nonumber
\label{hh1} \\
&&\times \sum_{j=0}^{N-1}\Phi _j\left( \frac \xi {C_{\alpha -1}}\right) \Phi
_j\left( \frac \eta {C_{\alpha -1}}\right) ,  \label{hh1} \\
H_{\alpha <\beta }\left( \xi ,\eta \right) &=&\prod_{k=\alpha }^{\beta
-1}\left( \phi _k\sqrt{\pi }\right) 
\mathop{\rm e}
{}^{F_\alpha \left( \xi \right) -F_\beta \left( \eta \right) }  \nonumber
\label{hh2} \\
&&\times \sum_{j=0}^{N-1}\frac{\Phi _j\left( \frac \xi {C_{\alpha -1}}%
\right) \Phi _j\left( \frac \eta {C_{\beta -1}}\right) }{\left[
1+\sum_{k=\alpha }^{\beta -1}\phi _k^2\right] ^{j/2}},  \label{hh2} \\
H_{\alpha >\beta }\left( \xi ,\eta \right) &=&\frac 1{\prod_{k=\beta
}^{\alpha -1}\left( \phi _k\sqrt{\pi }\right) }%
\mathop{\rm e}
{}^{F_\alpha \left( \xi \right) -F_\beta \left( \eta \right) }  \nonumber
\label{hh3} \\
&&\times \sum_{j=0}^{N-1}\frac{\Phi _j\left( \frac \xi {C_{\alpha -1}}%
\right) \Phi _j\left( \frac \eta {C_{\beta -1}}\right) }{\left[
1+\sum_{k=\beta }^{\alpha -1}\phi _k^2\right] ^{-j/2}}.  \label{hh3}
\end{eqnarray}
\end{mathletters}
The second term in Eq. (\ref{k1}) is found from Eqs. (\ref{k3}) and (\ref{wv}%
), 
\begin{eqnarray}
E_{\alpha ,\beta }\left( \xi ,\eta \right) &=&\frac{\prod_{k=\alpha }^{\beta
-1}\left( \phi _k\sqrt{\pi }\right) 
\mathop{\rm e}
{}^{G_\alpha \left( \xi \right) -G_\beta \left( \eta \right) }}{\sqrt{\pi
\sum_{k=\alpha }^{\beta -1}\phi _k^2}}  \nonumber  \label{e-full} \\
&&\times \exp \left\{ -\frac{\left( \xi -\eta \right) ^2}{\sum_{k=\alpha
}^{\beta -1}\phi _k^2}\right\}  \label{e-full}
\end{eqnarray}
for $\beta \geq \alpha +2$, while $E_{\alpha ,\alpha }=0$ and $E_{\alpha
,\alpha +1}=w_\alpha $. Here the function $G_\alpha $ reads 
\begin{equation}
{}^{}G_\alpha \left( \xi \right) =\frac{\xi ^2}2\left( \phi _\alpha
^{-2}-\phi _{\alpha -1}^{-2}\right) .  \label{gg}
\end{equation}

Now, we are in position to compute the matrix kernels $K_{\alpha ,\beta }$
via Eqs. (\ref{k1}), (\ref{hh}) and (\ref{e-full}) in the leading order in $%
N\rightarrow \infty $ and keeping $X_k=\phi _kN^{1/2}/\pi \sim {\cal O}%
\left( 1\right) $ fixed. The simplest, diagonal kernel $K_{\alpha ,\alpha }$
can be evaluated through the Christoffel--Darboux formula \cite{Szego-1967},
supplemented by the asymptotics of Hermite functions, 
\begin{equation}
\left\{ 
\begin{array}{c}
\Phi _{2N}\left( t\right)  \\ 
\Phi _{2N+1}\left( t\right) 
\end{array}
\right\} \simeq \frac{\left( -1\right) ^N}{N^{1/4}\sqrt{\pi }}\left\{ 
\begin{array}{c}
\cos \left( 2tN^{1/2}\right)  \\ 
\sin \left( 2tN^{1/2}\right) 
\end{array}
\right\}   \label{herfun}
\end{equation}
where $t\sim \Delta {\cal O}\left( N^0\right) $. One obtains, 
\begin{equation}
K_{\alpha ,\alpha }\left( \xi ,\eta \right) =%
\mathop{\rm e}
{}^{G_\alpha \left( \xi \right) -G_\alpha \left( \eta \right) }\frac{\sin
\left[ \pi \Delta ^{-1}\left( \xi -\eta \right) \right] }{\pi \Delta
^{-1}\left( \xi -\eta \right) }.  \label{k-diag}
\end{equation}
Two other cases, $\alpha <\beta $ and $\alpha >\beta $, demand more effort.
For $\alpha <\beta $ we represent the sum for $H_{\alpha <\beta }$ in Eq. (%
\ref{hh2}) as a difference of two series, $\sum_{j=0}^\infty \left( \ldots
\right) -\sum_{j=N}^\infty \left( \ldots \right) $. The first sum is exactly
computable by making use of the Mehler summation formula \cite{Szego-1967}.
In the thermodynamic limit, this procedure yields a term which is equal to $%
E_{\alpha ,\beta }$ in Eq. (\ref{e-full}), and therefore it gets canceled
from the expression (\ref{k1}) for $K_{\alpha <\beta }$ which is completely
due to the remaining sum $\sum_{j=N}^\infty \left( \ldots \right) $. To
evaluate the latter, we replace the sum over $j$ by an integral to get 
\begin{eqnarray}
&&K_{\alpha <\beta }\left( \xi ,\eta \right) =-\prod_{k=\alpha }^{\beta
-1}\left( \phi _k\sqrt{\pi }\right) 
\mathop{\rm e}
{}^{G_\alpha \left( \xi \right) -G_\beta \left( \eta \right) }  \nonumber \\
&&\times \int_1^\infty d\lambda _1\cos \{\pi \frac{\xi -\eta }\Delta \lambda
_1\}\exp \{-\frac{\pi ^2\lambda _1^2}2\sum_{k=\alpha }^{\beta -1}X_k^2\}.
\label{alb}
\end{eqnarray}
In the case $\alpha >\beta $ the large--$j$ terms in Eq. (\ref{hh3}) yield
the main contribution to the sum due to the factor $[1+\sum_{k=\beta
}^{\alpha -1}\phi _k^2]^{j/2}$. Then, passing from summation to integration,
we derive 
\begin{eqnarray}
&&K_{\alpha >\beta }^{}\left( \xi ,\eta \right) =\frac 1{\prod_{k=\beta
}^{\alpha -1}\left( \phi _k\sqrt{\pi }\right) }%
\mathop{\rm e}
{}^{G_\alpha \left( \xi \right) -G_\beta \left( \eta \right) }  \nonumber \\
&&\times \int_0^1d\lambda \cos \{\pi \frac{\xi -\eta }\Delta \lambda \}\exp
\{\frac{\pi ^2\lambda ^2}2\sum_{k=\beta }^{\alpha -1}X_k^2\}.  \label{agb}
\end{eqnarray}
Notice that the structure of the block matrix in Eq. (\ref{det}) allows one
to simultaneously suppress the prefactors of the form $\prod_k\left( \ldots
\right) 
\mathop{\rm e}
^{\left( \ldots \right) }$ in Eqs. (\ref{k-diag}), (\ref{alb}) and (\ref{agb}%
). Having this in mind, we come down to the closed analytical determinantal
expression Eq. (\ref{det}) for $(p_0+\ldots +p_d)$--point density--density
correlator with $K_{\alpha ,\beta }$ replaced by $M_{\alpha ,\beta }$, 
\begin{mathletters}
\label{m-full}
\begin{eqnarray}
M_{\alpha ,\alpha }\left( \xi ,\eta \right)  &\equiv &\frac{\sin \left[ \pi
\Delta ^{-1}\left( \xi -\eta \right) \right] }{\pi \Delta ^{-1}\left( \xi
-\eta \right) },  \label{sin} \\
M_{\alpha <\beta }\left( \xi ,\eta \right)  &\equiv &-\int_1^\infty d\lambda
_1\cos \{\pi \frac{\xi -\eta }\Delta \lambda _1\}  \nonumber  \label{malb} \\
&&\times \exp \{-\frac{\pi ^2\lambda _1^2}2\sum_{k=\alpha }^{\beta
-1}X_k^2\},  \label{malb} \\
M_{\alpha >\beta }^{}\left( \xi ,\eta \right)  &\equiv &\int_0^1d\lambda
\cos \{\pi \frac{\xi -\eta }\Delta \lambda \}  \nonumber  \label{magbrg} \\
&&\times \exp \{\frac{\pi ^2\lambda ^2}2\sum_{k=\beta }^{\alpha -1}X_k^2\}.
\label{magb}
\end{eqnarray}
\end{mathletters}
Equations (\ref{det}) and (\ref{m-full}) are the main result of the paper.
They provide a detailed information about higher order parametric
density--density correlations in the case of multiparameter perturbation of
disordered system. Several particular correlators can be readily deduced
from our general expression: (i) For the scalar perturbation, one obtains
that $k_{p,q}=\Delta ^{p+q}\left\langle {}\right. \prod_{i=1}^p\nu (E+\omega
_i,0)\prod_{j=1}^q\nu (E+\Omega _j,\phi )\left. {}\right\rangle $ is
determined by 
\begin{equation}
k_{p,q} \equiv %
\mathop{\rm Det}
\left( 
\begin{array}{cc}
M_{\alpha ,\alpha }\left( \omega _i,\omega _j\right)  & M_{\alpha <\beta
}\left( \omega _i,\Omega _j\right)  \\ 
M_{\alpha >\beta }^{}\left( \Omega _i,\omega _j\right)  & M_{\alpha ,\alpha
}\left( \Omega _i,\Omega _j\right) 
\end{array}
\right) ,  \label{scalar}
\end{equation}
where $M_{\alpha ,\beta }$ are those given by Eqs. (\ref{m-full}) with $%
\sum_kX_k^2\rightarrow X^2$; (ii) By replacement \cite{SLA-1993} $\omega
_i/\Delta \rightarrow r_i$, $\Omega _i/\Delta \rightarrow R_i$ and $%
X^2\rightarrow -2i\tau $ in Eq. (\ref{scalar}) one arrives at the
space--time correlation function in the CSM model with a coupling $\lambda =1
$; here the coordinates $\left\{ r_i\right\} $ correspond to the time $t=0$,
while the $\left\{ R_i\right\} $ refer to the time $t=\tau $.

In summary, we presented a random--matrix--theory treatment of the problem
of higher order parametric spectral statistics in disordered systems with
broken time reveral symmetry in the presence of the multiparameter
perturbation. A complete analytical solution was based on the mapping the
initial problem onto a model of random Hermitian matrices coupled in a
chain. As a particular case of the general solution given by Eqs. (\ref{det}) 
and (\ref{m-full}), the multipoint
parametric spectral correlator Eq. (\ref{scalar}) for the scalar perturbation has been
obtained. Together with a well established correspondence between CSM
fermions and parametric level statistics, the latter expression provides an
information about the space--time correlation function in the
Calogero--Sutherland--Moser model of free, non--interacting fermions.

The authors thank I. Yurkevich for bringing the reference \cite{IIKS-1990}
to our attention.

\widetext


\begin{references}
\bibitem{SzA-1993}  A. Szafer and B. L. Altshuler, Phys. Rev. Lett. {\bf 70}%
, 587 (1993).

\bibitem{SA-1993}  B. D. Simons and B. L. Altshuler, Phys. Rev. Lett. {\bf 70%
}, 4063 (1993); B. D. Simons and B. L. Altshuler, Phys. Rev. B {\bf 48},
5422 (1993).

\bibitem{SLA-1993}  B. D. Simons, P. A. Lee, and B. L. Altshuler, Phys. Rev.
Lett. {\bf 70}, 4122 (1993).

\bibitem{M-1991}  M. L. Mehta, {\it Random Matrices} (Academic Press,
Boston, 1991).

\bibitem{E-1983}  K. B. Efetov, Adv. Phys. {\bf 32}, 53 (1983).

\bibitem{VWZ-1985}  J. J. M. Verbaarschot, H. A. Weidenm\"{u}ller, and M. R.
Zirnbauer, Phys. Rep. {\bf 129}, 367 (1985).

\bibitem{NS-1993}  O. Narayan and B. S. Shastry, Phys. Rev. Lett. {\bf 71},
2106 (1993).

\bibitem{BR-1994}  C. W. J. Beenakker, Phys. Rev. Lett. {\bf 70}, 4126
(1993); C. W. J. Beenakker and B. Rejaei, Physica A {\bf 203}, 61 (1994).

\bibitem{BK-1994}  M. V. Berry and J. Keating, J. Phys. A {\bf 27}, 6167
(1994).

\bibitem{ALM-1997}  A. M. Ozorio de Almeida, C. H. Lewenkopf, and E. R.
Mucciolo, Los Alamos preprint archive, chao--dyn/9711017.

\bibitem{SLA-1994}  B. D. Simons, P. A. Lee, and B. L. Altshuler, Phys. Rev.
Lett. {\bf 72}, 64 (1994).

\bibitem{TSA-1995}  N. Taniguchi, B. S. Shastry, and B. L. Altshuler, Phys.
Rev. Lett. {\bf 75}, 3724 (1995).

\bibitem{ZH-1995}  M. R. Zirnbauer and F. D. M. Haldane, Phys. Rev. B {\bf 52%
}, 8729 (1995), and references therein.

\bibitem{NP}  B. D. Simons, P. A. Lee, and B. L. Altshuler, Nucl. Phys. B 
{\bf 409}, 487 (1993); B. L. Altshuler and B. D. Simons, in: {\it Mesoscopic
Quantum Physics}, edited by E. Akkermans et al., Proceedings of Les Houches
Session LXI 1994 (Elsevier, 1995).

\bibitem{IIKS-1990}  A. R. Its, A. G. Izergin, V. E. Korepin, and N. A.
Slavnov, Int. J. Mod. Phys. {\bf 4}, 1003 (1990).

\bibitem{M-1997}  B. Eynard and M. L. Mehta, J. Phys. A {\bf 31}, 4449
(1998); G. Mahoux, M. L. Mehta, and J.--M. Normand, J. Phys. A {\bf 31},
4457 (1998); M. L. Mehta and P. Shukla, J. Phys. A {\bf 27}, 7793 (1994); M.
L. Mehta, Comm. Math. Phys. {\bf 79}, 327 (1981).

\bibitem{RemarkP}  The relation $\phi _k=\pi N^{-1/2}X_k$ can be obtained,
for instance, by comparison of supersymmetric generating functionals derived
for the model of coupled Hermitian random matrices with that \cite{SA-1993}
known for the microscopic problem of electron motion in disordered potential
in the $0D$ limit of nonlinear sigma--model \cite{E-1983,VWZ-1985}.

\bibitem{IZ-1980}  C. Itzykson and J. B. Zuber, J. Math. Phys. {\bf 21}, 411
(1980).

\bibitem{RemarkM}  More precisely, one should write $E+\omega _{i_\sigma
}^{\left( \sigma \right) }$ instead of $\omega _{i_\sigma }^{\left( \sigma
\right) }$. As far as we will be interested in the local regime, in which $%
K_{\alpha ,\beta }\left( E,E^{\prime }\right) =K_{\alpha ,\beta }\left(
E-E^{\prime }\right) $ is translationally invariant, one may omit $E$ in
both arguments.

\bibitem{RemarkH}  The Hermite functions are fixed by the orthogonality
relation $\int dt\exp \left[ -t^2\right] H_j\left( t\right) H_k\left(
t\right) =\delta _{jk}$ for the associated Hermite polynomials.

\bibitem{Szego-1967}  G. Szeg\"{o}, {\it Orthogonal Polynomials} (American
Mathematical Society, Providence, 1967).
\end{references}
\end{document}